\documentclass[11pt,oneside,english,a4paper]{elsart}
\usepackage[T1]{fontenc}
\usepackage[latin9]{inputenc}
\usepackage{float}
\usepackage{textcomp}
\usepackage{relsize}
\usepackage{graphicx}
\usepackage{amssymb}

\makeatletter

\DeclareRobustCommand{\lyxmathsym}[1]{\ifmmode\begingroup\def\b@ld{bold}
  \def\rmorbf##1{\ifx\math@version\b@ld\textbf{##1}\else\textrm{##1}\fi}
  \mathchoice{\hbox{\rmorbf{#1}}}{\hbox{\rmorbf{#1}}}
  {\hbox{\smaller[2]\rmorbf{#1}}}{\hbox{\smaller[3]\rmorbf{#1}}}
  \endgroup\else#1\fi}

\providecommand{\tabularnewline}{\\}

\makeatother

\usepackage{babel}

\begin{document}
\begin{frontmatter}

\title{\textrm{\huge Universality in short-time critical gluodynamics with
heat-bath- inspired algorithms}}

\author{Rafael B. Frigori}

\ead{frigori@utfpr.edu.br}

\address{Universidade Tecnológica Federal do Paraná, \\
Rua XV de Novembro 2191, CEP 85902-040, Toledo (PR), Brazil}
\begin{abstract}
Short-time dynamics technique is used to study the relaxation process
for the (2+1)-dimensional critical gluodynamics of the SU(2) lattice
gauge theory. A generalized class of heat-bath-inspired updating algorithms
was employed during the short-time regime of the dynamic evolution
for performance comparison. The static and dynamic critical exponents
of the theory were measured, serving as a dynamic benchmark for algorithmic
efficiency. Our results are in agreement with predictions from universality
hypothesis and suggest that there is an underlying universal dynamics
shared by the analyzed algorithms. 
\end{abstract}
\end{frontmatter}

\section{Introduction}

The pioneering study on equilibrium critical phenomena addressed by
Fisher et. al. through finite-size-scaling relations \cite{Fisher}
was soon extended to include a description of dynamic relaxation processes
\cite{Halperin,Suzuki}. Thus, it was demonstrated that in many systems
an universal behavior holds even far from equilibrium \cite{Janssen,Janssen_A,Huse}.
In lattice-gauge theories, Monte Carlo relaxation dynamics was investigated
by Okano et. al. in seminal works \cite{OkanoSchulkeYamagishiZheng,OkanoSchulkeZheng,Otobe_Lat04,Otobe_IJMPC}
where critical-exponents were computed. These studies reinforced the
hypothesis that universality among spin-systems and gauge-theories
\cite{Svet&Yaffe} is dynamically realized. In addition, predictions
from universality were checked in lattice-gauge theories beyond the
determination of critical-exponents. For instance, it was confirmed
that, near the deconfinement phase-transition, the dynamically generated
spectrum of gluonic-screening masses obeys universal ratios \cite{Fiore,Falcone,Frigori}.

Equilibrium Monte Carlo simulations allow for non-perturbative numerical
calculations, so they are essential for the study of some fundamental
theoretical issues on phase-transitions of lattice-gauge theories
\cite{GeneralGroups}. However, when approaching the thermodynamical
limit, those simulations have their thermalization efficiency increasingly
afflicted by the so called critical-slowing-down effect \cite{Sokal},
which makes it costly to compute independent gauge configurations.
A possible way to circunvent that effect comes from Short-time dynamic
techniques \cite{OkanoSchulkeZheng,Otobe_IJMPC}, which enables critical
gluodynamics to be efficiently exploited by local \cite{Creutz,KennedyPendleton,CabibboMarinari}
or global \cite{Jaster_HMC} updating algorithms. Therefore, to devise
improved algorithms \cite{OHB,HadronsIX,OuroPreto} by better understanding
their equilibration features --- e.g. (over)relaxation dynamics ---
may be worthy for many applications.

In this context, we extend our previous comparative study over a class
of improved heat-bath-inspired thermalization algorithms \cite{HadronsIX,OuroPreto}
to the non-equilibrium regime, by using a (2+1)-dimensional SU(2)
lattice-gauge theory at critical temperature. It is known that dynamic-relaxation
exponents $\theta$ and $z$ --- namely in contradistinction to static
ones: $\alpha,\beta,\gamma,\delta,\nu$ --- are generally dependent
on the very dynamics of thermalization; so, they would serve as discriminants
for algorithmic classes. The article is organized as follows: in Section
2, general properties of short-time critical dynamics are reviewed;
an overview of usual heat-bath algorithm and its generalizations is
the theme of Section 3; the setup of our simulations and the description
of a new procedure to implement sharp initial states on lattice-gauge
theories is given in Section 4; general data analysis and simulation
results are provided on Section 5; we summarize our findings in Section
6.

\section{Short-time critical dynamics}

Renormalization group techniques allowed Janssen \textit{et al} \cite{Janssen}
to show that under suitable initial conditions some magnetic systems,
after being suddenly quenched to critical-temperature $T_{c}$, present
universal-scaling behavior even at early evolution. This phenomenon
is observed just after a microscopic transient time-scale $t_{\mu}$
has elapsed and lasts for a macroscopic relaxation-period $t_{macro}$
before thermalization is established. 

During the short-time process of relaxation, when the system is driven
to equilibrium by a Monte Carlo stochastic-dynamics, just mild finite-size-effects
\cite{OkanoSchulkeYamagishiZheng} are noticeable. Then, reliable
information can be extracted from the resulting time-series by considering
some generalized-scaling relations for usual observables. For instance
\cite{OkanoSchulkeYamagishiZheng,OkanoSchulkeZheng}, the $kth$ moment
of magnetization obeys \begin{equation}
M^{\left(k\right)}\left(t,\tau,L,m_{0}\right)=b^{-k\beta/\nu}M^{\left(k\right)}\left(b^{-z}t,b^{1/\nu}\tau,b^{-1}L,b^{x_{0}}m_{0}\right),\label{GeneralScalling}\end{equation}
where $m_{0}$ is the initial magnetization, $b$ is an arbitrary
spatial scaling factor, $L$ is the system size, $t$ is the time-evolution
parameter, and the reduced-temperature is $\tau=\left(T-T_{c}\right)/T_{c}$.
While $\beta$ and $\nu$ are the well-known critical-exponents of
equilibrium (i.e. \textit{static}), $z$ and $x_{0}$ are relaxation-dependent
(i.e \textit{dynamical}) exponents, and both groups label universality
classes. 

Considering a hot initial-state with sharply-defined (and small) magnetization
$m_{0}$, it was shown \cite{Janssen,Janssen_A} that for a quenched
system, some simpler power-law relations hold for the magnetization
\begin{equation}
M\left(t,m_{0}\right)\approx m_{0}t^{\theta},\label{M_t}\end{equation}
and its higher-moment

\begin{equation}
M^{\left(2\right)}\left(t\right)\sim t^{\left(d-2\beta/\nu\right)/z},\label{M2_t}\end{equation}
where $\theta=\left(x_{0}-\beta/\nu\right)/z$ and $d$ denotes the
spatial dimensionality. Analogously, the temporal-autocorrelation
$A\left(t\right)$ for the magnetization evolves as\begin{equation}
A\left(t\right)\sim t^{\theta-d/z}.\label{A_t}\end{equation}

Consequently, the critical phenomenology of a system can be studied
by observing the temporal-scaling behavior --- up to the time-scale
$t_{0}\sim m_{0}^{-z/x_{0}}$ \cite{Janssen} --- of some of its observables,
during the early relaxation (i.e. short-time regime) of the dynamic
evolution. This technique allows for the determination of equilibrium
and non-equilibrium critical-exponents \cite{Janssen} without severe
critical-slowing-down restrictions \cite{OkanoSchulkeYamagishiZheng}.
Hence, averages in the dynamic approach are taken over independent
samples, without relying on ergodical time-averaged measurements \cite{OkanoSchulkeZheng}.

\section{Heat-Bath inspired algorithms}

Thermalization is the process responsible for producing independent
configurations during Monte Carlo simulations. When the quenched approximation
is used in gauge theories, by setting the fermion determinant to unit,
it is possible to apply local algorithms such as Metropolis or heat-bath
(HB) \cite{Creutz,KennedyPendleton} as efficient first-choices for
thermalization. However, when a critical-point is approached, as in
the continuum limit of the theory, simulations suffer from critical-slowing-down
phenomenon \cite{Adler}. This drastically increases correlations
among successive field-configurations, which produces integrated auto-correlation
times $\tau_{int}$ raising as a power of the lattice side $L$. Thereby,
it induces Monte Carlo statistical errors to diverge as $\mathcal{O}\left(\sqrt{2\tau_{int}}\right)$
\cite{Sokal}.

In order to circumvent the critical-slowing-down, the standard HB-algorithm
and micro-canonical updates are combined, which allows for improved
generation of independent samples \cite{Adler}. In particular, in
the hybrid version of overrelaxed algorithms, $m$ energy-conserving
microcanonical update sweeps are done after a standard local ergodic
update.

For a $SU(2)$ lattice-gauge theory, the action can be factorized
as a sum of single-link actions $S_{1-link}$, which may be written
down as

\begin{equation}
S_{1-link}=-\frac{\beta}{2}Tr\left[U_{\mu}\left(x\right)H_{\mu}\left(x\right)\right],\label{S_1link}\end{equation}
where $U_{\mu}\left(x\right)\in SU\left(2\right)$ is a gauge-link,
and $H_{\mu}\left(x\right)$ --- the effective-magnetic field, written
as a sum over staples --- is proportional to an SU(2) matrix. As a
useful notation consider $H_{\mu}\left(x\right)=N_{\mu}\left(x\right)\tilde{H_{\mu}}\left(x\right)$
with $\tilde{H_{\mu}}\left(x\right)\in SU\left(2\right)$ and $N_{\mu}\left(x\right)=\sqrt{\det H_{\mu}\left(x\right)}$.

Then, by using Eq.(\ref{S_1link}) and the invariance of the group
measure under group multiplication, the HB update is obtained\begin{equation}
U_{\mu}^{old}\left(x\right)\longrightarrow U_{\mu}^{new}\left(x\right)=V\tilde{H_{\mu}^{\dagger}}\left(x\right),\label{HB_update}\end{equation}
where $V=v_{0}I+i\vec{\cdot v}\cdot\vec{\sigma}\in SU\left(2\right)$
is randomly generated by choosing $v_{0}$ according to the distribution
$\sqrt{1-v_{0}\lyxmathsym{\texttwosuperior}}$$\exp\left(\beta Nv_{0}\right)dv_{0}$
and $\vec{v}$ pointing along a uniformly chosen random direction
in $\mathbb{R}^{3}$, under the constraint $|V|^{2}=v_{0}^{2}+\vec{v}\vec{\cdot v}=1$. 

We have proposed in \cite{HadronsIX,OuroPreto} a modified HB algorithm
(MHB) in which the generation of the updating matrix $V$ is carried
out as usual, except for the additional step
\begin{itemize}
\item \textit{Transform the new vector-components of V as $\vec{v}\rightarrow-sgn(\vec{v}\cdot\vec{w})\vec{v}$,}
\end{itemize}
where $W=w_{0}I+i\cdot\vec{w}\cdot\vec{\sigma}=U_{\mu}^{old}\left(x\right)\tilde{H_{\mu}}\left(x\right)$,
and $sgn$ is the sign function.  

This may be also thought as an ergodical modification of the overheat-bath
(OH), an algorithm devised some years ago \cite{OHB}. The underlying
idea in both cases is to incorporate a micro-canonical move \cite{Adler}
into the heat-bath step. The difference is that while in MHB the vector
$\vec{v}$ is randomly set --- except for its sign, which is determined
according to the aforementioned rule --- in the OH algorithm, one
deterministically sets $\vec{v}=-\vec{w}$ --- without obeying a uniform
distribution for $\vec{v}$ --- and renormalizes it as $|\vec{v}|=\sqrt{1-v_{0}^{2}}$.
Therefore, OH incorporates a micro-canonical move in an \textit{exact}
algorithm, but it would not be ergodic \cite{OuroPreto}.

Despite being slightly more computer-consuming than OH, our modification
implements a micro-canonical move that explicitly preserves ergodicity
and allows for a reduction of about 20\% in statistical errors, at
the same computational cost, when compared with the usual HB algorithm
\cite{HadronsIX,OuroPreto}.

\section{Simulation setup}

Lattice gauge theories in d-dimensions can be easily turned to finite-temperature
formalism \cite{OkanoSchulkeZheng,Fiore,Falcone,Mendes_O4} by constraining
the euclidean spacetime volume to $V=L_{s}^{d-1}L_{t}$ --- under
the assumption $L_{t}\ll L_{s}$ --- where $L_{s}$ and $L_{t}$ are
the spatial and temporal lattice sides. The equilibrium temperature
is given by $T{}^{-1}=a\cdot L_{t}$, where $a$ is the physical lattice
spacing.

The Polyakov loop $P_{\vec{x}}$ on site $\vec{x}$ is a useful quantity
whose spatially averaged trace $\sum_{\vec{x}}Tr(P_{\vec{x}})$ is
the order parameter of the deconfinement phase-transition. It is defined
as an ordered product of gauge-links $U_{0}\left(x_{0},\vec{x}\right)$
in the temporal direction. Here we use, for the particular $SU\left(2\right)$
case, a non-equilibrium time-dependent definition\begin{equation}
L_{\vec{x}}\left[t\right]=Tr\left\{ P_{\vec{x}}\right\} \left[t\right]=Tr\left\{ \frac{1}{2}\prod_{x_{0=0}}^{L_{t}-1}U_{0}\left(x_{0},\vec{x}\right)\right\} \left[t\right],\label{Polyakov}\end{equation}
 where $\left[t\right]$ denotes a given instant in the Monte Carlo
time.

A strict analogy with the magnetization in Eq.(\ref{M_t}) is addressed
to gauge theories \cite{OkanoSchulkeZheng} by defining a time-dependent
\textit{effective magnetization} $M$ as\begin{equation}
M\left(t\right)\doteq\left\langle \frac{1}{L_{s}^{d-1}}\sum_{\vec{x}}L_{\vec{x}}\left[t\right]\right\rangle _{sample},\label{M_Poly}\end{equation}
where $\left\langle \cdots\right\rangle _{sample}$ stands for an
average taken over simultaneous samples, at an instant $t$. Similarly,
Eq.(\ref{M2_t}) is associated to the second-moment of $M$ by the
relation\begin{equation}
M^{\left(2\right)}\left(t\right)\doteq\left\langle \left(\frac{1}{L_{s}^{d-1}}\sum_{\vec{x}}L_{\vec{x}}\left[t\right]\right)^{2}\right\rangle _{sample},\label{M2_Poly}\end{equation}
 and Eq.(\ref{A_t}) is related to\begin{equation}
A\left(t\right)\doteq\left\langle \left(\frac{1}{L_{s}^{d-1}}\right)^{2}\sum_{\vec{x}}L_{\vec{x}}\left[t\right]L_{\vec{x}}\left[0\right]\right\rangle _{sample}.\label{A_Poly}\end{equation}

For the purpose of performing short-time dynamic simulations, it is
necessary to prepare hot-initial states $(T=\infty)$ with a sharply-defined
(small) effective magnetization $(M\approx m_{0})$. This may be achieved
by the procedure

\section*{Setting a sharp initial-state }
\begin{enumerate}
\item \textit{Restart the random-number generator using a new seed.} 
\item \textit{Set all components of each $SU(2)$ gauge-link randomly}. 
\item \textit{For a maximum number $N_{steps}$ of loops do} 
\item \textit{Evaluate $M\left(0\right)$ as in} Eq.(\ref{M_Poly})\textit{.
If $\left\Vert M\left(0\right)-m_{0}\right\Vert <\delta m_{tolerance}$
start} step-8\textit{, else do} 
\item \textit{Choose up to $N_{sites}$ lattice-sites at random positions
$\left(x_{0},\vec{x}\right)$. Then, update their temporal gauge-links
by $U_{0}\left(x_{0},\vec{x}\right)\rightarrow-sgn\left[M\left(0\right)-m_{0}\right]P_{\vec{x}}^{\dagger}U_{0}\left(x_{0},\vec{x}\right)$.
Where $P_{\vec{x}}$ is as in} Eq.(\ref{Polyakov}). 
\item \textit{Return to} step-4. 
\item \textit{Return to} step-3. 
\item \textit{Perform thermalization.} 
\end{enumerate}
In our simulation-setup, parameters were tuned to \textit{$N_{steps}\simeq\mathcal{O}\left(10^{4}\right)$}
and \textit{$N_{sites}\simeq0.2\times L_{s}^{d-1},$} which allowed
us to shortly obtain \textit{$\delta m_{tolerance}/m_{0}\simeq\mathcal{O}\left(10^{-3}\right),$}
without introducing any spurious spatiotemporal correlations among
samples.

\section{Numerical results}

We have used lattice sizes up to $128^{2}\times2$ in our simulations.
Initial lattice configurations were set using the previously described
algorithm. Despite the dynamic critical exponents in Eq.(\ref{M_t}),
Eq.(\ref{M2_t}), and Eq.(\ref{A_t}) being rigorously defined just
for $m_{0}=0$, the minimal initial magnetization we considered was
set to be $m_{0}=\left[400(1)\right]\cdot10^{-5}$ --- i.e., no extrapolation
to $m\rightarrow0$ was attempted --- which allowed us to keep better
signal-to-noise ratios.

To thermalize the gauge-fields, we employed the aforementioned heat-bath-like
algorithms and the standard Wilson action at critical-coupling $\beta=3.4505$
\cite{Otobe_IJMPC}. We have prepared 50000 independent initial samples
for simulations, whose measured observables {[}Eqs.(\ref{M_Poly})
--- (\ref{A_Poly}){]} were grouped in 10 blocks --- to estimate errors
--- for each lattice sweep. The Monte Carlo temporal evolution was
followed up to 500 steps, as seen in Figure(1): Left panels.

The best-fit range for each observable was determinated by searching
for the largest plateaus in the time-interval $\left(t_{\mu},500\right],$
within the minimal $\chi^{2}/dof$, which were produced as outcomes
for previous power-law fits, Figure(1): Right panels. Within the established
stable-regions we accurately fit the corresponding critical exponents.
This methodology also allowed us to unveil the thermalization efficiency
of the employed algorithms by their direct comparison during relaxation.

It is interesting to note that average numerical-values from data
of time-series produced by MHB-evolution are nicely located between
the ones obtained by HB and OH algorithms, Figure(1): Left panels.
Also, despite their different inner-dynamics, the algorithms we implemented
reproduced self-consistently --- up to good numerical approximation
--- the same set of static and dynamic critical-exponents, as seen
in Table(1).

The numerical results for critical exponents agree with previous studies
\cite{OkanoSchulkeZheng,Otobe_Lat04,Otobe_IJMPC,Jaster_HMC} that
investigated the universality-hypothesis \cite{Svet&Yaffe} between
SU(2) lattice gauge theory and the 2d-Ising model. Also, upon closer
examination for similarities among the time-series produced by different
algorithms --- and their dynamic $\theta$ and $z$ exponents ---
there is a strong indication that all heat-bath-inspired algorithms
analyzed here share the same universal non-equilibrium-relaxation
dynamics%
\footnote{This may be a hint that effects of some hypothetical \textit{ergodicity
violations,} previously conjectured in \cite{OuroPreto}, would be
less harmful than what is expected for some simulations \cite{Fiore}
using the OH algorithm. %
}.

\begin{figure}[H]
\centering\includegraphics[clip,width=8cm,height=6cm]{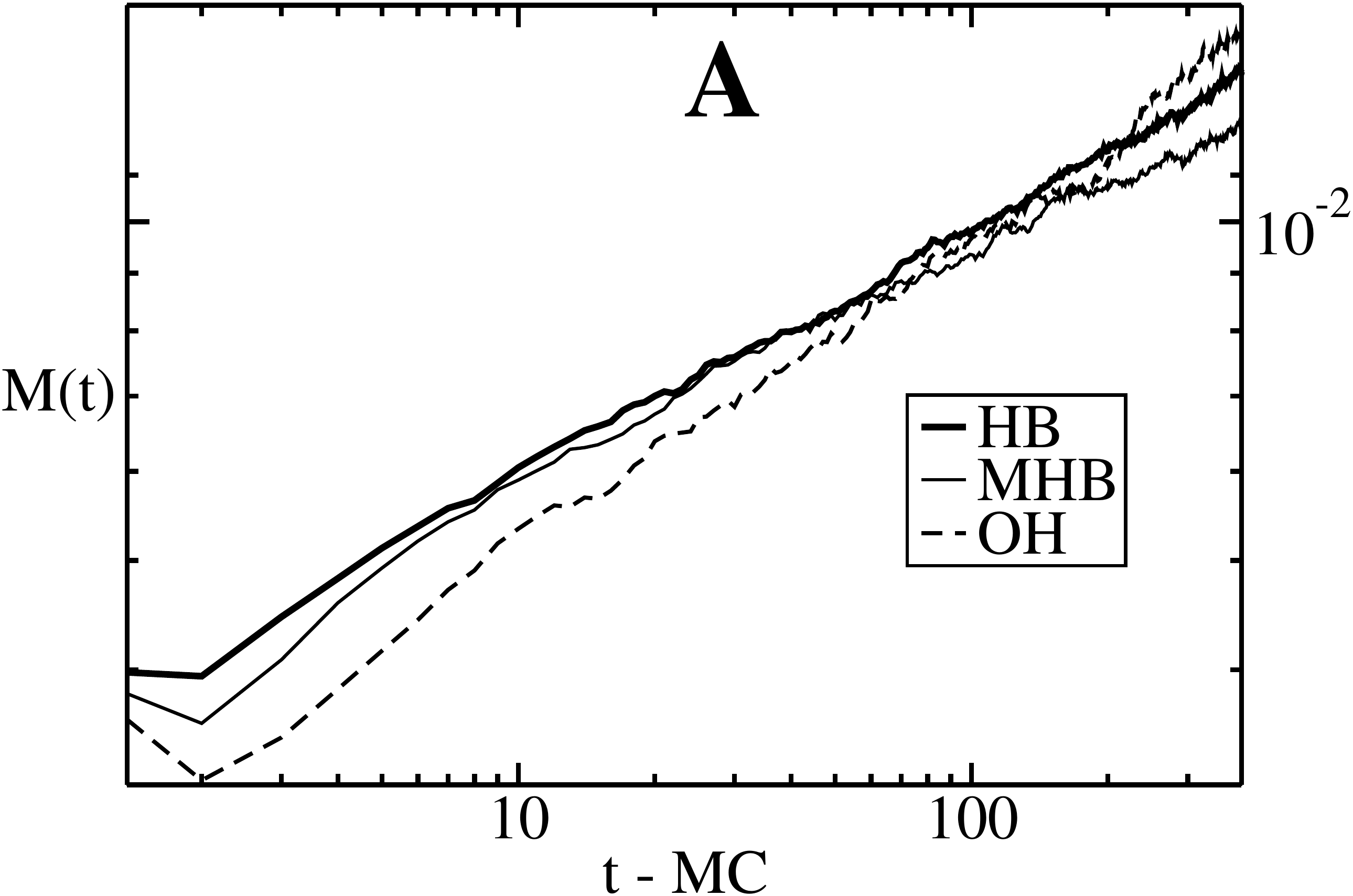}\includegraphics[clip,width=8cm,height=6cm]{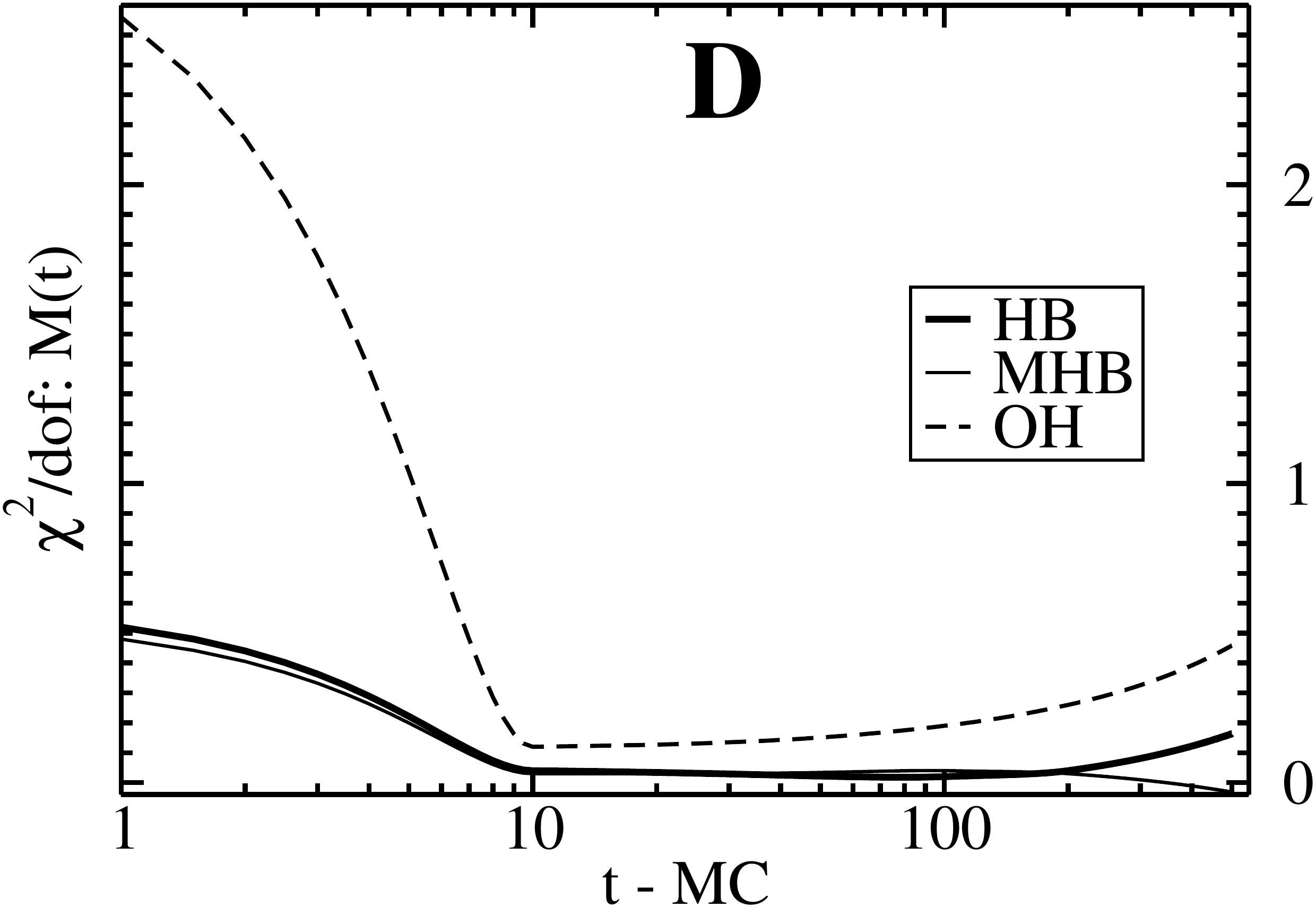}

\includegraphics[clip,width=8cm,height=6cm]{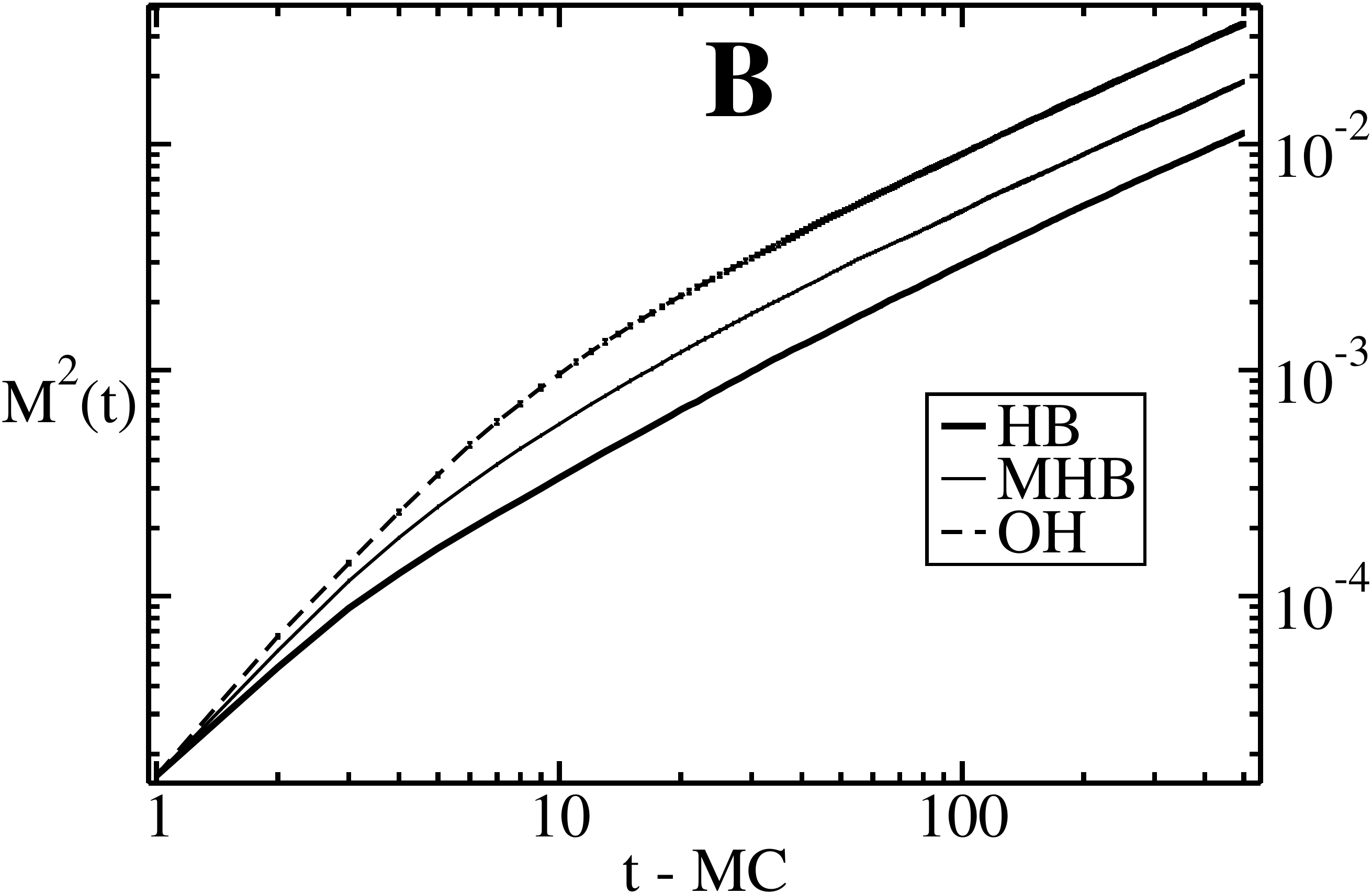}~~~\includegraphics[clip,width=8cm,height=6cm]{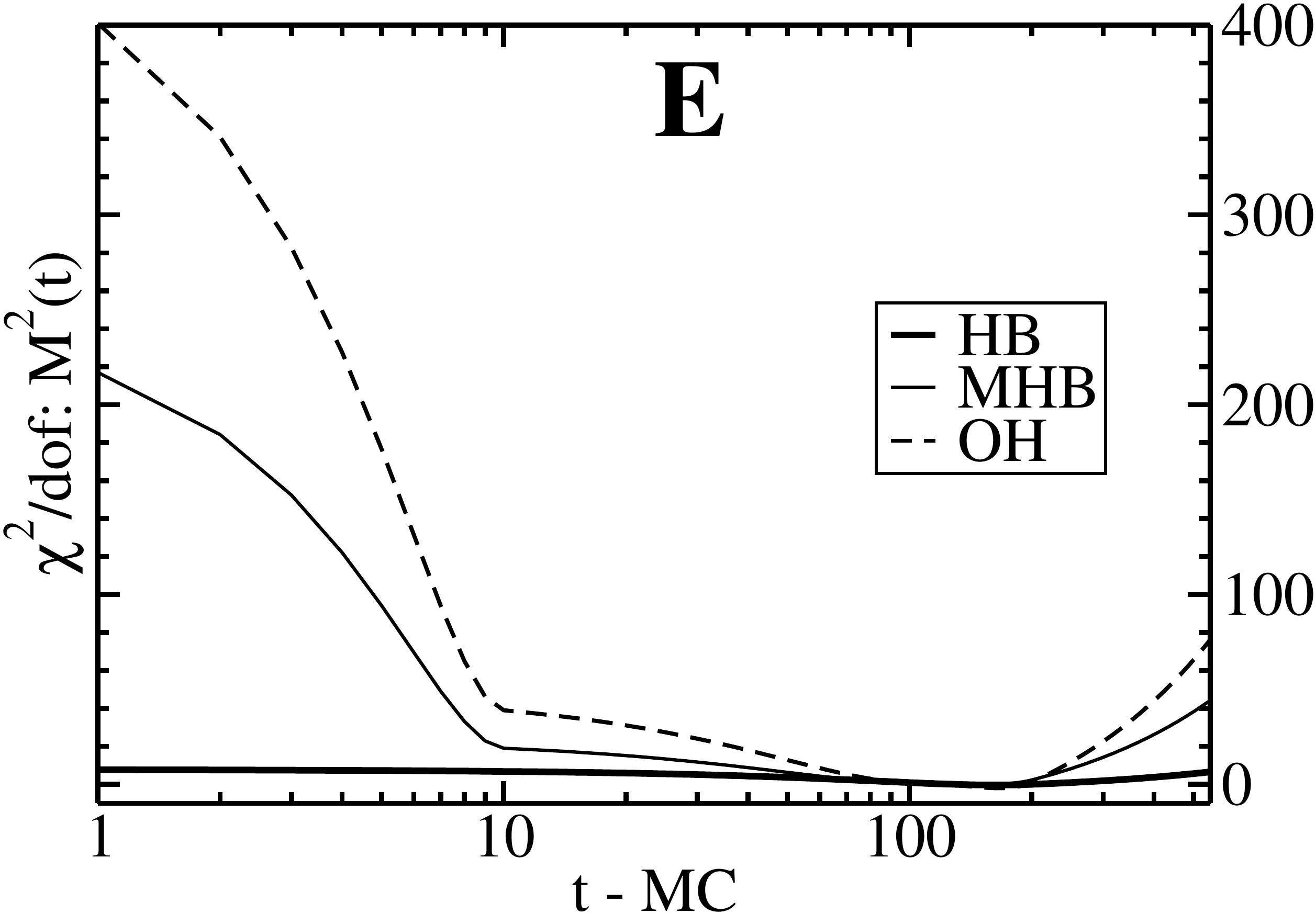}

\includegraphics[clip,width=8cm,height=6cm]{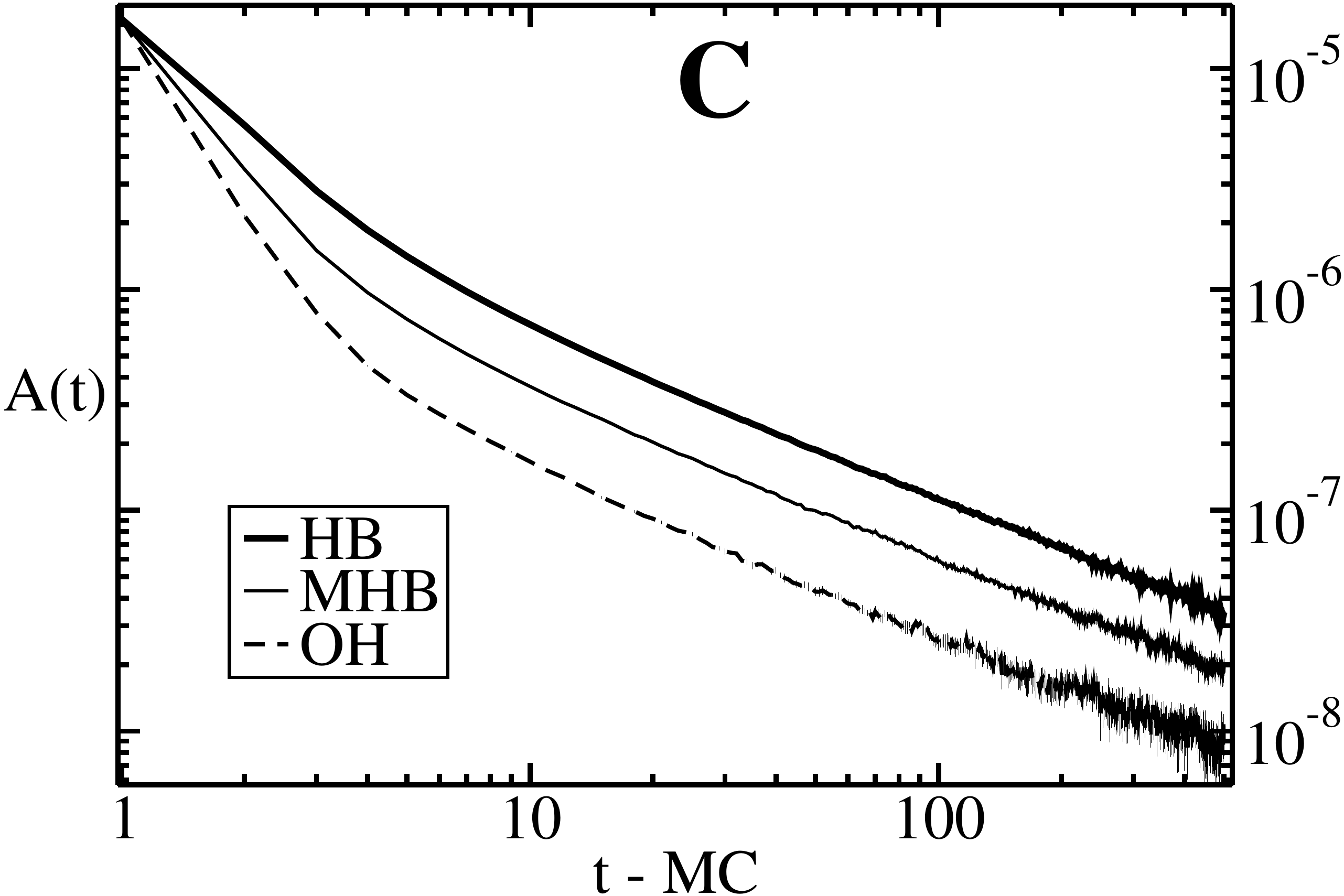}~~~\includegraphics[clip,width=8cm,height=6cm]{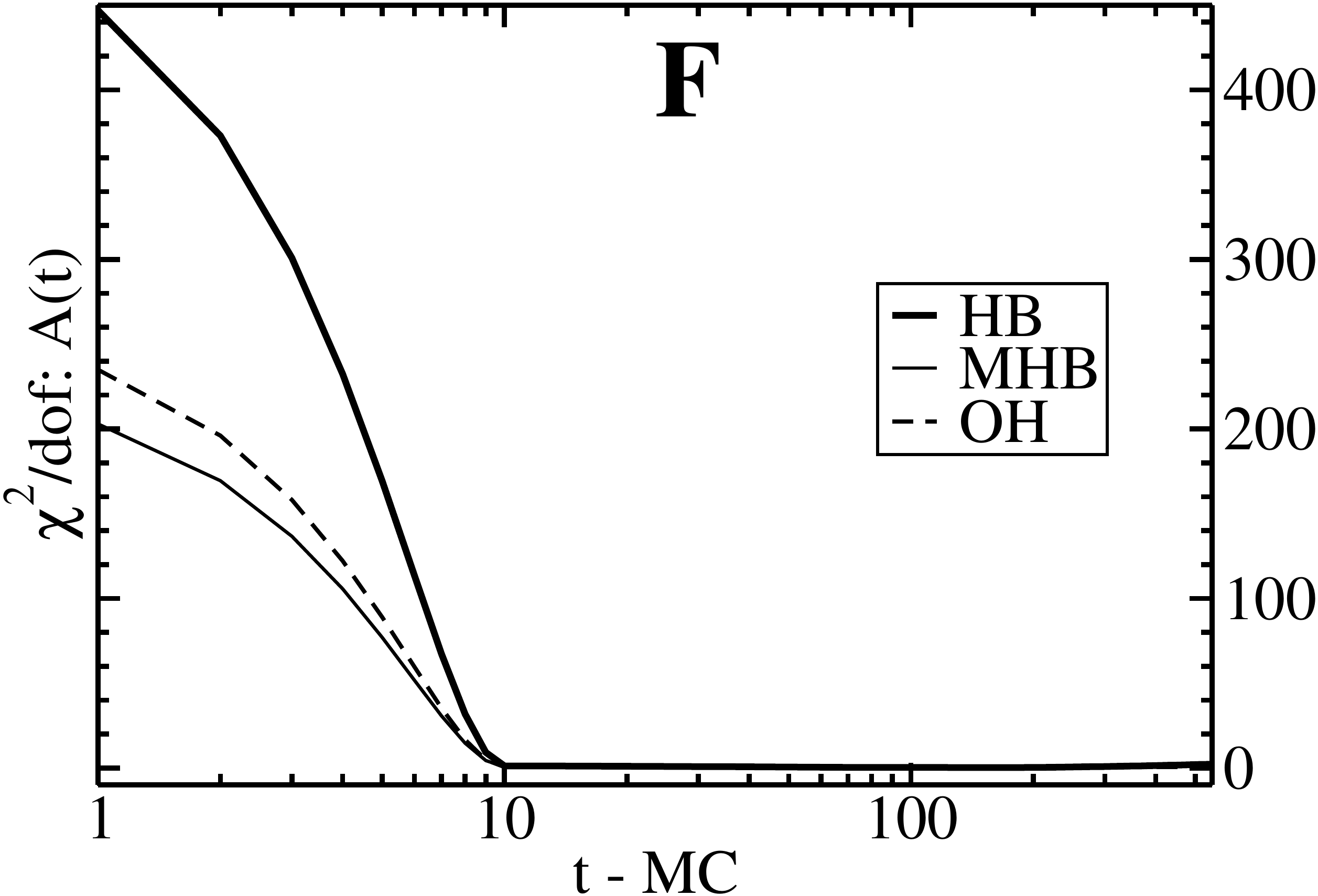}

\caption{Monte Carlo time-evolution of observables in short-time relaxation
dynamics for different algorithms. (\textbf{Panel A}) and (\textbf{Panel
B}) are plots for the first and second moments of the magnetization,
(\textbf{Panel C}) is the temporal auto-correlation of the magnetization.
Error bars are simple Monte Carlo standard deviations, they have been
omitted (on Panel A) for better visualization.\protect \\
Detection of the fitting-range $\left[t_{\mu},t\right]$ by analysis
of the $\chi^{2}/dof$ vs. $t$ produced as an outcome from power-law
fit anzates for different observables. (\textbf{Panel D}) and (\textbf{Panel
E}) are plots for the fit-quality for the first and second moments
of the magnetization, (\textbf{Panel F}) is the fit-quality for the
temporal auto-correlation of the magnetization. }

\end{figure}

\begin{table}[H]
\centering\begin{tabular}{|l|c|c|c|}
\hline 
Algorithm  & $\theta$  & $z$  & \multicolumn{1}{c|}{$\beta/\nu$}\tabularnewline
\hline 
HB  & 0.1976(5)  & 2.108(3)  & 0.110(1)\tabularnewline
\hline 
MHB  & 0.1843(7)  & 2.141(4)  & 0.114(5)\tabularnewline
\hline 
OH  & 0.1955(1)  & 2.094(7)  & 0.132(7)\tabularnewline
\hline 
Ising-2d (HB)  & 0.191(1)  & 2.155(3)  & 1/8\tabularnewline
\hline
\end{tabular}

\caption{Static and dynamic critical-exponents obtained with each algorithm
employed for time-evolution in our Monte Carlo simulations. The last
line presents the same quantities obtained for the two-dimensional
Ising model \cite{Otobe_IJMPC}.}

\end{table}

\section{Summary}

Our results extend previous studies on short-time dynamics of lattice
gauge theories \cite{OkanoSchulkeZheng,Otobe_Lat04,Otobe_IJMPC,Jaster_HMC}
to a whole class of heat-bath-inspired thermalization algorithms.
This would be seem as a self-consistent cross-check of the universality
hypothesis \cite{Svet&Yaffe} holding far-from-equilibrium. Here,
this is verified for a set of thermalization algorithms that embraces
different relaxation dynamics.

Notwithstanding the particularities of each overrelaxation-dynamics
embedded in the algorithms we analyzed, they seem to display an underlying
universal dynamics, a fact that is also corroborated by the numerical
values obtained for the dynamical critical-exponents. Thus, in this
comparative study for benchmarking thermalization algorithms, when
focusing on finding a better balance among most desirable features
for updating algorithms --- e.g., faster decorrelation of samples
and higher signal-to-noise levels at same computer cost --- it is
observed that MHB presents the most desirable algorithmic realization.

The methods employed in this article can also be easily adapted to
dynamic analysis of gauge theories for larger unitary, symplectic,
or exceptional gauge groups \cite{GeneralGroups} or even to $O\left(N\right)$
continuous-spin systems \cite{OuroPreto}. That may bring some new
perspectives for non-equilibrium comparative studies on universality
in full-QCD \cite{Mendes_O4}. Also, some general constraints for
the infrared gluon-propagator --- written in close similarity with
magnetization-moments \cite{GluonConstraint} --- would be investigated
by short-time dynamic simulations, though with ameliorated finite-size
and critical-slowing-down effects.

\section*{Acknowledgments}

The author thanks Tereza Mendes and Attilio Cucchieri for useful discussions
during the early stage of this research. Financial support was provided
by FAPESP and CAPES (Brazil).

\end{document}